\documentclass[floatfix, noshowpacs, preprintnumbers, twocolumn, amsmath, amssymb, aps, prl, superscriptaddress]{revtex4}

\pdfoutput=1

\usepackage{graphicx, xcolor}
\usepackage{soul}
\usepackage{slashed}
\usepackage{bm} \usepackage{graphicx} \usepackage{amsmath}
\usepackage{amsthm,stmaryrd}
\usepackage{amssymb} \usepackage{physics} \usepackage{txfonts}
\usepackage{color}
\usepackage{comment}
\usepackage{ulem}
\usepackage{appendix}
\usepackage{booktabs} 
\usepackage{float}
\usepackage{threeparttable}
\usepackage{diagbox}

\begin{document}

\title{Quantum support vector machines for aerodynamic classification}

\author{Xi-Jun Yuan}
\altaffiliation{These authors contributed equally to this work.}
\affiliation{Center for Integrated Quantum Information Technologies (IQIT), School of Physics and Astronomy and State Key Laboratory of Advanced Optical Communication Systems and Networks, Shanghai Jiao Tong University, Shanghai 200240, China}

\author{Zi-Qiao Chen}
\altaffiliation{These authors contributed equally to this work.}
\affiliation{School of Mechanical Engineering, Key Lab of Education
Ministry for Power Machinery and Engineering, Gas Turbine Research
Institute, Shanghai Jiao Tong University, Shanghai 200240, China}

\author{Yu-Dan Liu}
\affiliation{School of Mechanical Engineering, Key Lab of Education
Ministry for Power Machinery and Engineering, Gas Turbine Research
Institute, Shanghai Jiao Tong University, Shanghai 200240, China}

\author{Zhe Xie}
\affiliation{Center for Integrated Quantum Information Technologies (IQIT), School of Physics and Astronomy and State Key Laboratory of Advanced Optical Communication Systems and Networks, Shanghai Jiao Tong University, Shanghai 200240, China}

\author{Xian-Min Jin }\email{xianmin.jin@sjtu.edu.cn}
\affiliation{Center for Integrated Quantum Information Technologies (IQIT), School of Physics and Astronomy and State Key Laboratory of Advanced Optical Communication Systems and Networks, Shanghai Jiao Tong University, Shanghai 200240, China}
\affiliation{TuringQ Co., Ltd., Shanghai, 100190, China}
\email{xianmin.jin@sjtu.edu.cn}

\author{Ying-Zheng Liu}
\affiliation{School of Mechanical Engineering, Key Lab of Education
Ministry for Power Machinery and Engineering, Gas Turbine Research
Institute, Shanghai Jiao Tong University, Shanghai 200240, China}

\author{Xin Wen }\email{wenxin84@sjtu.edu.cn}
\affiliation{School of Mechanical Engineering, Key Lab of Education
Ministry for Power Machinery and Engineering, Gas Turbine Research
Institute, Shanghai Jiao Tong University, Shanghai 200240, China}

\author{Hao Tang }\email{htang2015@sjtu.edu.cn}
\affiliation{Center for Integrated Quantum Information Technologies (IQIT), School of Physics and Astronomy and State Key Laboratory of Advanced Optical Communication Systems and Networks, Shanghai Jiao Tong University, Shanghai 200240, China}

\begin{abstract}
{Aerodynamics plays an important role in aviation industry and aircraft design. Detecting and minimizing the phenomenon of flow separation from scattered pressure data on airfoil is critical for ensuring stable and efficient aviation. However, since it is challenging to understand the mechanics of flow field separation, the aerodynamic parameters are emphasized for the identification and control of flow separation. It has been investigated extensively using traditional algorithms and machine learning methods such as the support vector machine (SVM) models. Recently, a growing interest in quantum computing and its applications among wide research communities sheds light upon the use of quantum techniques to solve aerodynamic problems. In this paper, we apply qSVM, a quantum SVM algorithm based on the quantum annealing model, to identify whether there is flow separation, with their performance in comparison to the widely-used classical SVM. We show that our approach outperforms the classical SVM with an 11.1\% increase of the accuracy, from 0.818 to 0.909, for this binary classification task. We further develop multi-class qSVMs based on one-against-all algorithm. We apply it to classify multiple types of the attack angles to the wings, where the advantage over the classical multi-class counterpart is maintained with an accuracy increased from 0.67 to 0.79, by 17.9\%. Our work demonstrates a useful quantum technique for classifying flow separation scenarios, and may promote rich investigations for quantum computing applications in fluid dynamics. }
\end{abstract}

\maketitle

Detecting and minimizing the phenomenon of flow separation is of critical meaning for ensuring stable and efficient aviation. The flow separation means the boundary layer is no longer able to remain attached to the airfoil and detaches from the surface.  Airfoil is the main lifting surface of the aircraft and increasing the angle of attack will allow it to deflect more fluid and produce a greater lift. However, once the angle of attack reaches a certain critical value, a sudden decrease in the lift force happens and the aircraft will stall. As demonstrated in Fig. 1, the wake created behind the separation region affects the pressure distribution on the airfoil, significantly reducing lift and increasing drag. Aircraft stall is a serious issue for aviation, leading to aerodynamic loss, maneuverability decrease and even the crush of airplane. Thus, detecting flow separation is extensively researched by academics and engineers in the past decades\cite{Tung2001,Siauw2009,Ghouila-Houri2018,Zhou2021}, using various methods such as shear stress sensors \cite{Tung2001}, distributed pressure signals\cite{Siauw2009} , or compatible calorimetric micro-sensors \cite{Ghouila-Houri2018}. Yet, detecting the flow separation based on distributed pressure sensors on the airfoil is still challenging since flow separation is an intricate phenomenon affected by multiple factors such as incoming flow velocity, pressure and surrounding flow field. Therefore, further work on identification and control of flow separation by aerodynamic parameters is worth investigating.\par

Machine learning (ML) has flourished in fluid mechanics recently for its ability to extract informative features from data. It provides a powerful information processing framework that can be used for understanding, modeling, optimizing, and controlling fluid flows. ML is widely used in the flow reconstruction. The velocity fields can be faithfully predicted based on scattered pressure sensors and local measurements using Ensemble Kalman filter, artificial neural networks and dynamic mode decomposition with time delay \cite{Le Provost2021,Manohar2022,Yuann2021}. Missing data in the flow measurement can also be reconstructed based on data fusion and deep generative adversarial model\cite{Wen2019, Buzzicotti2021}. The flow control parameter optimization and performance prediction can be achieved using support vector regression, model-free reinforcement learning and a deep Koopman embedding \cite{Lee2022, Jiao2021, Mendible21}. Classification is a classical problem for ML, which is also widely used in fluid dynamics to distinguish between various flow conditions and dynamic regimes. Support vector machines (SVMs) is one of the typical ML algorithms for classifying problem. It is a supervised learning method developed on the basis of statistical learning theory, embodying the structural risk minimization principle. They are gaining popularity due to the features such as simple structure, global optimality and strong robustness. Ling $\&$ Templeton \cite{Ling2015} used SVMs to classify and predict regions of high uncertainty in the Reynolds stress tensor and compared the results with Adaboost decision trees and random forests. Wang $et~al.$ \cite{Wang2015} examined the feasibility of applying SVMs to high angle-of-attack unsteady aerodynamic modeling of an airfoil. Shi $et~al.$\cite{Shi2018} applied SVMs to optimize spacecraft multidisciplinary design by using filter-based sequential radial basis function. In spite of the advantages above, SVMs are unsuited to implement for large-scale training samples due to enormous storage and calculation cost. Moreover, the general need for multi-class classification in real applications raise even more demand of the computation resources.

\begin{figure*}[t!]
    \centering
	\includegraphics[scale = 0.78]{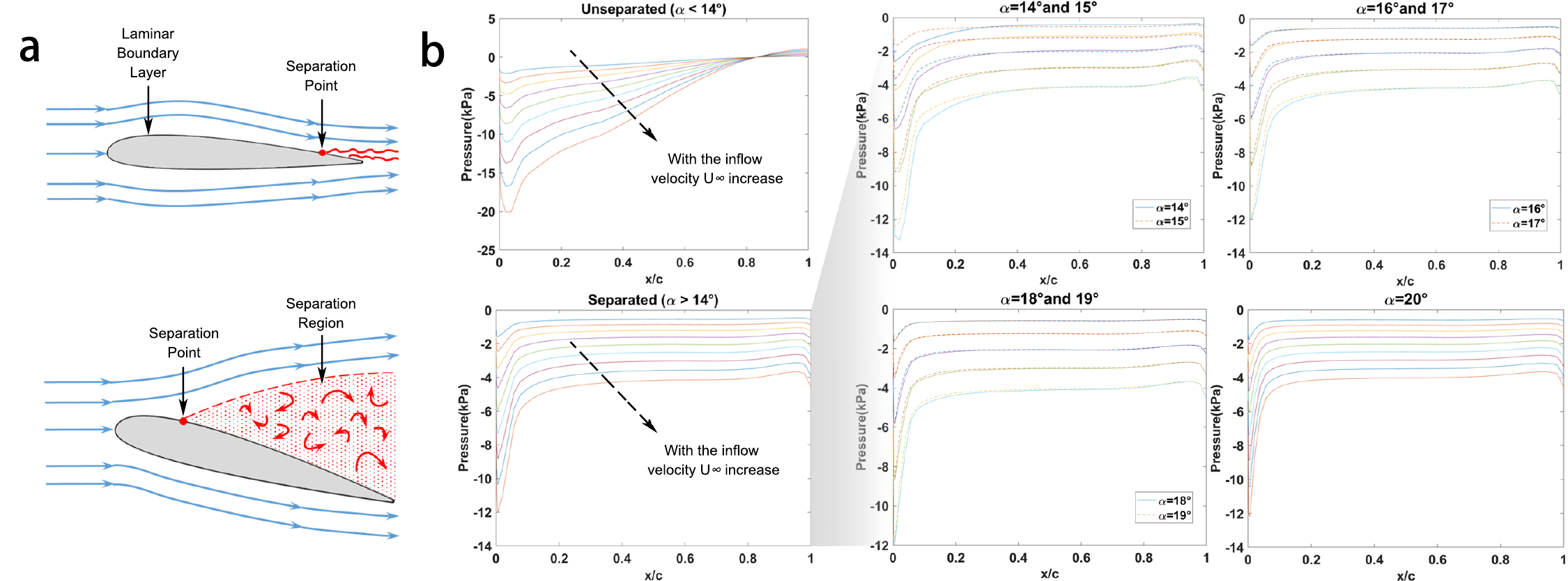}
	\caption{Illustration of the flow separation and pressure distribution on the airfoil. (a) shows the scenarios of flow without separation and flow separation, respectively. There exists a steep velocity gradient in the boundary layer, generated by the air viscosity. The velocity near the surface is close to zero while the flow maintains a high speed in the distance due to the inertial force. As the flow velocity increases, the air at the end of the wing will have a negative velocity due to the reverse pressure gradient and begin to generate flow separation. (b) depicts the trend of pressure distribution without and with flow separation, respectively. The pressure serves as a physical principle for detecting flow separation on the airfoil. $\alpha$ here denotes the AOA (angle of attack), which represents the angle between the forward direction of the wing and the chord. Before stall, the sudden change of pressure distribution from steep slope to flat plateau indicates that flow separation occurs. While after the flow separation, the trend of the pressure distribution does not change significantly with further increasing of angle of attack, but simply shifts to a lower level. The critical AOA is 14 $^\circ$ in this case, which means greater AOAs will lead to the separation. The flow separation was divided into 4 groups (14-15 $^\circ$,16-17 $^\circ$,18-19 $^\circ$,20 $^\circ$) and manifests the homologous pressure distribution trend, which are meant to be distinguished by the multi-class qSVMs we proposed in this paper later. }
	 
\end{figure*}

Now those tasks may be more efficiently addressed with the aid of quantum computing. As an alternative way of encoding data and setting algorithms, quantum computing has demonstrated advantages over classical computers, and is being extensively studied for applications such as finance\cite{Tang2020b}, biology\cite{Jin2022}, computer vision\cite{Shi2020}, $etc$. There are also a few emerging theoretical works\cite{Cao2013, Arrazola2019, Childs2020, Engel2019, Wang2020, Gaitan2020} on quantum application in computational fluid dynamics using quantum amplitude estimation (QAE)\cite{Gaitan2020} and quantum Fourier transform (QFT). It has been proven that setting ML components in the quantum way, such as quantum kernels and quantum feature maps, can lead to general advantages that make the rationale for the emerging field of quantum ML\cite{Huang2021,Schuld2015,Wang2022}, and hence a few quantum-circuit-based SVM algorithms were proposed\cite{Rebentrost2014,Schuld2018, Havlicek2019}. 

Still, it is worth noting that current quantum technologies remain to be noisy and of an intermediate scale. This sets challenges on the scalability for those pure quantum gate models such as QAE, QFT and the aforementioned quantum SVM. Apart from gate models, quantum annealing, which is currently more scalable with large analog evolution paths\cite{Kadowaki1998,Crosson2021} , has also been applied to ML tasks\cite{Mott2017,Li2018,Davis2021, Willsch2020}. Recently, quantum annealing has been employed to implement SVM for large binary classification tasks\cite{Willsch2020}. Quantum annealing 
utilize quantum tunnelling effects that assist the hop out of a local minimum\cite{Kadowaki1998}. Meanwhile, since quantum annealing usually generates an ensemble of different close-to-global-minimum solutions due to the physical feature of diabatic annealing, it captures broader and unseen data than the single global minimum\cite{Willsch2020}, such yielding a more suitable decision on the hyperplane and hence better performance than classical SVM. So far, quantum classification techniques have never been applied to tasks in fluid mechanics. 

In this paper, we adopt a quantum-annealing-based SVM, qSVM, to detect whether there exists flow separation. We compare its performance with the widely-used classical SVM from Scikit, and show our approach outperforms the classical SVM for this binary classification task. The classical SVM shows an accuracy of 81.8\% while our qSVM reaches 90.9\%, with an increase of accuracy by 11.1\%. We further develop a multi-class qSVMs based on the one-versus-rest strategy. We apply it to detect the AOA of airfoil and exhibit the advantage over the classical multi-class counterpart at the same time. The qSVM shows an accuracy of 79\%, yielding an increase of accuracy by 17.9\% comparing to the classical accuracy of 67.0\%. Therefore, we for the first time, to the best of our knowledge, apply quantum algorithms to classification tasks in aerodynamics, which may greatly promote further and deeper explorations for rich quantum computing applications in aerodynamics and fluid field analysis.

\section{Method}
\subsection{Map SVM to quadratic programming problems}
Support vector machine (SVM) is a supervised machine learning model for classification and regression, which can be used in aerodynamics analysis such as classifying flow separation, angle of attack, and inflow velocity, $etc$. Consider a dataset
\begin{equation}
		D = \{(x_n; t_n=-1\ or \ 1),n=0,\cdots, N-1\},
\end{equation}
where $D$ is a point set in d-dimensional space, $x_n\in R^d$ is the spatial coordinate of each point, and $t_n=\pm1$ denotes the target class assigned to $x_n$ in the classification problem. SVM selects a hyperplane in d-dimensional space to separate two point sets where the selection criterion is to maximize the distance between the two points set and the hyperplane. In some cases, the two points set are so mixed in their eigenspace that they can’t be separated by a hyperplane in d-dimensional space, the solution is to introduce kernel function $k(x_n,x_m)$ to map points into higher dimensional space and make the points linearly separable in that dimension. In subsequent sections, we choose the Gaussian kernel function due to its high applicability to high-dimensional problems. 

The solution for finding the hyperplane for SVM can be mapped to a quadratic programming (QP) problem which aims at minimizing the objective function $E$

\begin{eqnarray}
		&&minimize\qquad E=\frac{1}{2}\sum_{mn}\alpha_{m}\alpha_{n}t_{m}t_{n}k(x_m,x_n)-\sum_{n}\alpha_n,
	    \\&&sub\!ject\; to\qquad  0\leq \alpha_n \leq C,
	    \\&&\qquad \qquad \qquad\sum_{n}\alpha_n t_{n} = 0,
\end{eqnarray}
for $N$ coefficients $\alpha_n\in R$, where $C$ is a regularization parameter. The optimal solution to the equation $\alpha^\ast = {(\alpha_{1}^\ast,\alpha_{2}^\ast,\cdots,\alpha_{N}^\ast)}^T$ is used to construct the decision function 
\begin{equation}
	f(x)= \sum_{n} \alpha_{n} t_n k(x_n,x) +b , 
\end{equation}
for an arbitrary point $x \in R^d$. A reasonable choice of bias $b$ is \cite{press2007numerical}
\begin{equation}
	b = \frac{\sum_n \alpha_{n}(C-\alpha_n)[t_{n}-\sum_{m} \alpha_{m} t_{m} k(x_{m},x_{n})]}{\sum_{n} \alpha_{n}(C-\alpha_{n})}
\end{equation}
The form represents the positional relationship between the point and the hyperplane, where $f(x)>0$ indicates the point above the hyperplane, $f(x)<0$ indicates below, and $f(x)=0$ means the point lies on the hyperplane. Using the decision function we can predict the class of sample $x$ by $\widetilde{t} = sign(f(x))$. In the following, we symbolically write cSVM to denote classical SVM defined by Eqs. (2)-(4).

\subsection{Quantum annealing and qSVM}
We now introduce our quantum version of SVM, which is driven by quantum annealing. Quantum annealing (QA) is an algorithm that uses quantum fluctuations to obtain the lowest energy state, $i.e.$, the global optimal solution. It runs as following: The system is initially set to an arbitrary superposition, then we change the magnitude of the added Hamiltonian $H^{\prime}$ and the system begins quantum evolution according to the time-dependent $Schr\ddot{o}dinger \ equation$ 
\begin{equation}
		i\hbar \frac{\partial \psi}{\partial t}=[H_{0}+H^{\prime}(t)]\psi, 
\end{equation}
where the added Hamiltonian $H^{\prime}$ is changed and the resulting states begin to tunnel from one potential well to other adjacent potential well, during which process the ability of quantum parallel computing is demonstrated. Eventually, the perturbation is slowly removed and the system reaches the ground state corresponding to $H_0$. As an ideal consequence the desired global minimization is acquired and we get our best solution of the QP problem. 
\par The Quantum Annealing solver used in our paper is Advantage\_system4.1 from D-Wave Ocean SDK\cite{ref14}. At the end of the anneal, a single solution is sampled from a set of good solutions, with some probability, and returned. Thus it actually produces some close-to-optimal solutions($i.e.$, usually have a slightly higher energy than the global minimum) due to noises in hardware, but it still meets the requirement for classification. Since the D-Wave system that executes the QA algorithm can only perform discrete, binary calculations, the real number form QP problem needs to be transformed into a quadratic unconstrained binary optimization (QUBO) problem\cite{Willsch2020} (See details in Appendix A). It contains several hyperparameters: $B$ designates the base used for the encoding, $K$ is the number of binary variables to encode $\alpha_n$, $\xi$ is a multiplier in QUBO problem and $\gamma$ is the hyperparameter of the kernel function. After the QUBO formulation, we are ready to conduct the quantum version of SVM (qSVM) on the quantum annealer.

\subsection{Multiclass qSVM  }
We further introduce the one-against-all approach to construct the multi-class SVMs. In the one-against-all approach, the multi-class problem is decomposed into $k$ binary classification where $k$ is the number of classes of the multi-classification problem. Each binary classification is composed of one of the $k$ classes and the rest as the other class, and we set $t_n=1$ for points 
belonging to former class and $t_n=-1$ for points of the latter. Therefore, each classifier will determine whether the point $x$ belongs to one of the $k$ classes. Subsequent processing follows the method of the binary classification, as having been done in the previous sections, and repeat the procedure $k$ times. Then each point $x$ will have $k$ $f(x)$ values obtained by $k$ classifiers, and the class corresponding to the largest value among them stands for the class we predict.

In order to evaluate and understand the power of qSVM, we use both measurement data obtained from aerodynamic experiment and the simulated data via CFD (computational fluid dynamics) \cite{wilcox1998turbulence} simulation. Briefly, there are two classification tasks, deciding whether the flow is separated or not (binary classification), and judging the AOA of the airfoil after flow separation (multi-class classification). Field experiment was performed in an open-circuit wind tunnel. From the averaged velocity field obtained by PIV measurement, the flow separation phenomenon was visualized, and the corresponding relationships between the onset of flow separation and the variable parameters (angle of attack, freestream velocity) were identified. The angle of attack was adjusted in the range $\alpha$ = 0–19$^{\circ}$ by turning the rotary table with airfoil fixed on it. The average freestream velocity $U$ was maintained at 10, 13, and 17 m/s. 10 pressure taps were distributed in a section at the mid-span of the airfoil model to accurately detect the flow separation and there are totally 45 cases. CFD simulation provided more data, the angle of attack varied from 0-20$^{\circ}$ successively with a common difference of 1$^{\circ}$ and the freestream velocity raised from 40 m/s to 120 m/s, increasing 10 m/s each time. There were 106 pressure measuring points distributed uniformly over the whole airfoil surface. See more details on experimental setup in Appendix B.

\section{Result Analysis}
\subsection{Binary classification application}
The dataset we use in binary classification is  measurement data acquired from the aerodynamic experiment. The sample size is not very large and we regard it as a demonstration of small sample classification.
\par The dataset obtained by experiment consist of 45 points $(x_n,t_n)$, each $x_n$ is a $10 \times 1$ vector. A total of 10 pressure taps were nonlinearly distributed in a section at the mid-span of the airfoil model. The pressure values of these ten points are assigned to $x_1$ to $x_{10}$ in sequence. 27 points denotes negative and they stand for the flow without separated, others denoting positive stand for the flow separation. Since the pressure distribution under different aerodynamic parameters can be described in a space with much lower dimensionality \cite{Zhou2021}, we first compress the data using PCA(principal component analysis) and use the first two dimensions to represent the raw data, this can help make the result visualized and save computing resources. Then the dataset divides into two parts, 34 points for training and 11 points for test. For qSVM, we used the 
lowest energy solution returned by D-Wave Advantage\_system4.1. 

\begin{figure}[th]
\includegraphics[scale=0.4]{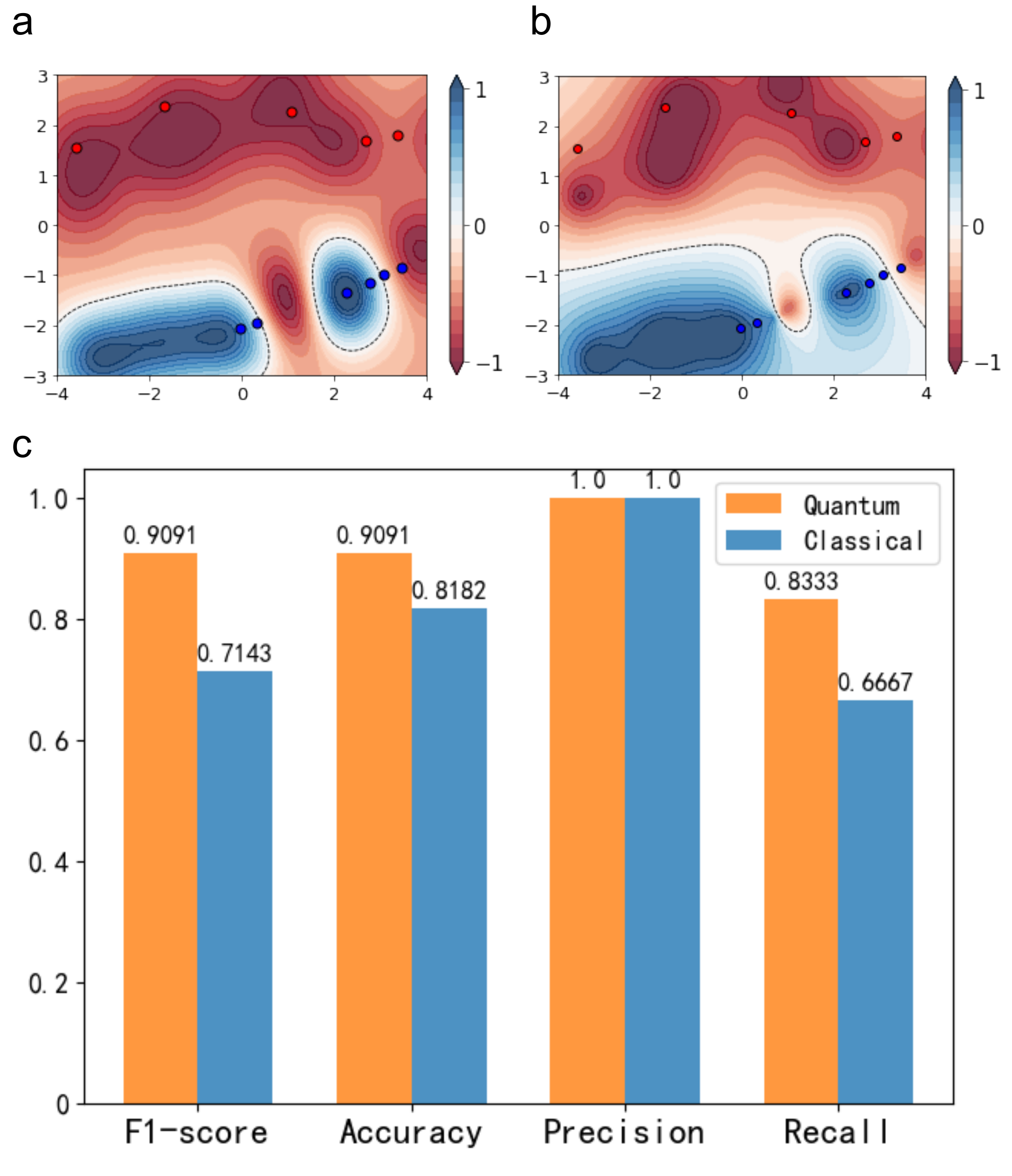} 
\caption{Results of flow classification problem. \quad (a) Decision plane obtained by cSVM and (b) qSVM for the same dataset, the classical classifier used here is svm.SVC supplied by scikit-learn 1.0.2. The hyperparameters for the SVMs are $B= K= 2, \xi = 0, \gamma = 1$, and $\  C = 3$. The color represents the value of the training data decision function $f(x)$. The black dotted line representing $f(x)=0$ marks the decision boundary. Test set data are represented by circles, with red representing real class $t_n = -1$ and blue representing real class $t_n= 1$, and the class of prediction is determined by their corresponding background color. (c) Comparing the performance of quantum and classical classifiers on flow separation tasks via different indicators. Refer to Appendix 
C for the meaning of these indicators.}
\end{figure}

To better compare the results of the classical and quantum classifiers, we use the same training and test sets for both classifiers, the results of airflow classification are demonstrated in Fig.2. We display the decision boundary by a curve separating two different color areas, the blue area represents $t_n = 1$ and the red area represents $t_n = -1$, dark colored areas represent the concentrated distribution of the training set data. Test set data are also drawn in the form of scattered points to visualize the predictive ability of the classifier.

 It can be seen that the decision boundary $f(x)=0$ of both classifiers can distinctly distinguish the regions where the training set data is clustered, and with the help of color gradient, we see that the decision boundary meets the requirements of SVM classification, that the boundary lies farthest from the closest training data points (the support vectors). More than 80\% test data are distributed in the color background corresponding to the real class, indicating that both classifiers can make effective judgments on whether the flow field is separated. Furthermore, Fig.2(c) elaborates that compared with cSVM, qSVM has achieved better performance in various indicators, including increasing the accuracy from 0.818 to 0.909 by 11.1\%. See Appendix C for detailed definition of binary classification indicators. 


\begin{figure*}[t!]
	\centering
	\includegraphics[scale = 0.7]{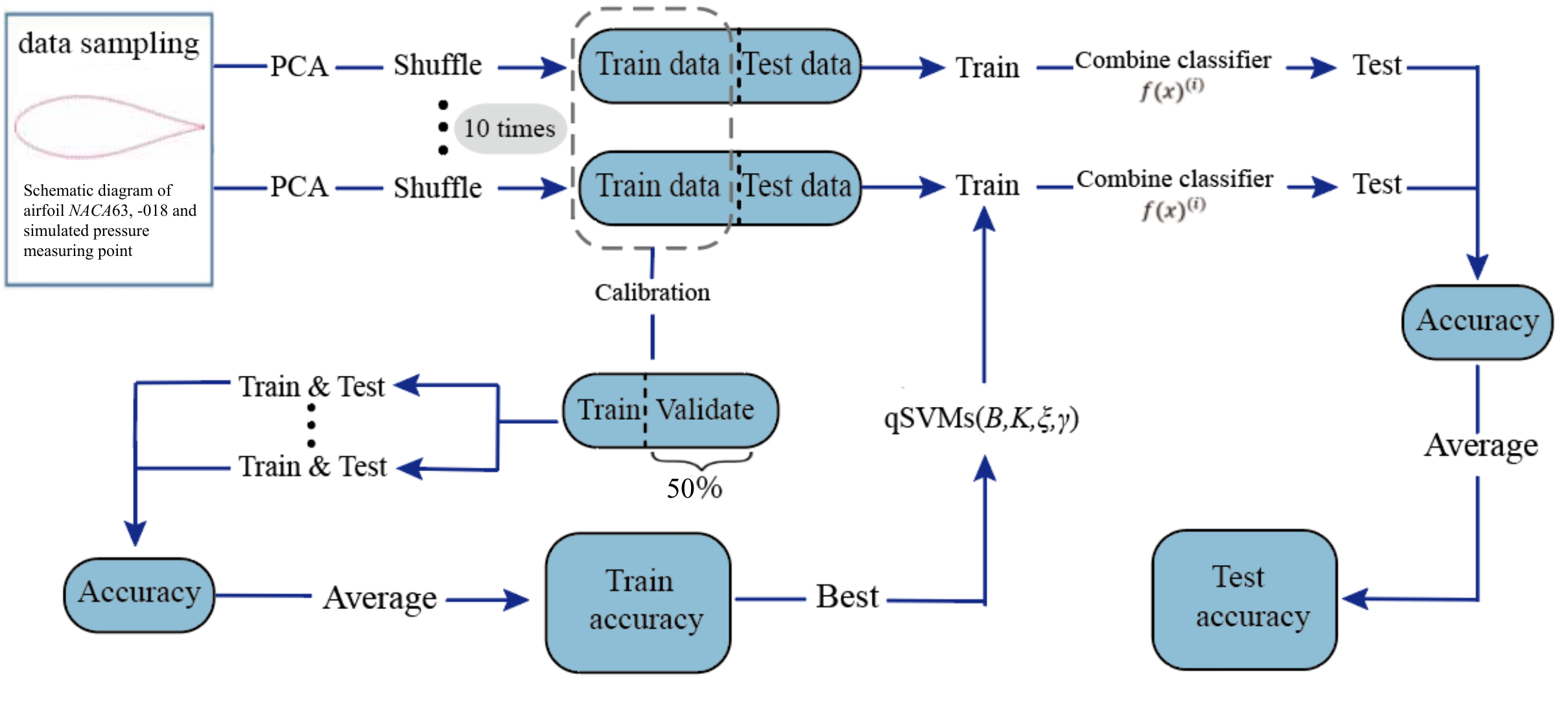}
	\caption{Process for handling AOA classification. For angle classification, the first step is data sampling. Data are obtained from 106 pressure sensors and each measurement forms a vector $x_n\in R^{106}$. Then the data is subjected to data dimensionality reduction through PCA, after which the vector reduces to $x_n\in R^{3}$. In the next step, the data is shuffled in different orders and divided into 10 sets of training and test data. Each training set is sent to SVMs to train and test its classification accuracy, and then the average of 10 sets of accuracy is obtained. Next, by changing the parameters to maximize the average accuracy, we got the best parameters for the corresponding problem. In the test phase, five lowest energy solutions returned by the D-Wave Advantage\_system4.1 are combined to form the classifier, the process is repeated 10 times and the classification accuracy is output. 
	 }
\end{figure*}

\subsection {Attack angles classification  }
As illustrated in Fig.1, predicting the angle of attack is a critical problem. When the AOA increases, the flow separation zone expands gradually and finally causes the stall. The lift decreases abruptly and the aircraft gets out of control. Therefore, classifying the angle of attack is an important task for aerodynamics. We implement the multi-class qSVM for this task, and compare it with the classical counterpart, both classifiers following the same one-against-all strategy.

\par As the angle of attack after separation is much more difficult to distinguish than that before flow separation, the dataset we chose is the flow field data after flow separation, which was obtained via CFD simulation. Actually, we also classified the angle of attack before fluid separation. The accuracy of classification is very close to 100\% for both cSVMs and qSVMs, and there is scarcely any difference in classification ability, so it is not shown in the article. The dataset consists of data with AOA from 14$^\circ$ to 20$^\circ$, including 9 data at each angle. We classify the adjacent two degrees as one class while the data of 20$^\circ$ as one class alone, dividing the data set into four classes. Each data is reduced to three dimensions by PCA method for the angle of attack of the flow field can be distinguishable in three dimensions.

\par  We then divided the data into training data and test data, including 43 points and 20 points in three-dimensional space respectively. In order to better test the classification ability, we shuffle the data ten times and get different training sets and test sets. Each training data is trained separately to select the hyperparameters that maximize the average accuracy. The entire process is shown in Fig.3.

\par  The multi-class qSVM used in this task consists of four binary qSVMs. For each classifier, five best solutions (labeled qSVM($B,K,\xi,\gamma,i$) for $i$ in range 0-4) are combined by averaging over the decision function to better exert the advantages of quantum annealing. This means the $\alpha_n^{(i)}$, $b^{(i)}$ corresponding to each optimal solution will construct a decision function $f(x)^{(i)}$ (see Eq.(5)), and the decision function $F(x)$ of each classifier is the average of $f(x)^{(i)}$. We then predict the class of a test point $x$ by finding the classifier corresponding to the maximum decision function value and give the homologous forecast class. 

\par After comparing the qSVM with traditional classifier which also uses one-against-all algorithm, we get the classification result shown as Fig. 4. It shows the classifier's performance of the dataset shuffled in five different orders, denoted as G1, G2, G3, G4 and G5. We see the width for blue-green sectors in Fig. 4a is much narrower than in Fig.4b, suggesting that for a case of dataset G1, qSVM shows a higher accuracy for classifying AOA than cSVM. 

\par We further show in Fig. 4c the quantitative results for different shuffled datasets ranging from G1 to G5, where qSVMs always get an accuracy no less than cSVMs. This shows that through the combination of several close-to-optimal solutions, qSVMs can distinguish different types of data more effectively. Since the prediction of the class in the multi-class problem is based on the value of the decision function $f(x)$ rather than its valuence, this advantage can therefore be clearly observed. In addition, we found that the two classifiers didn't have more serious classification errors outside the adjacent classes, which also shows the effectiveness of the classifiers.

\begin{figure}[th]
\includegraphics[scale=0.4]{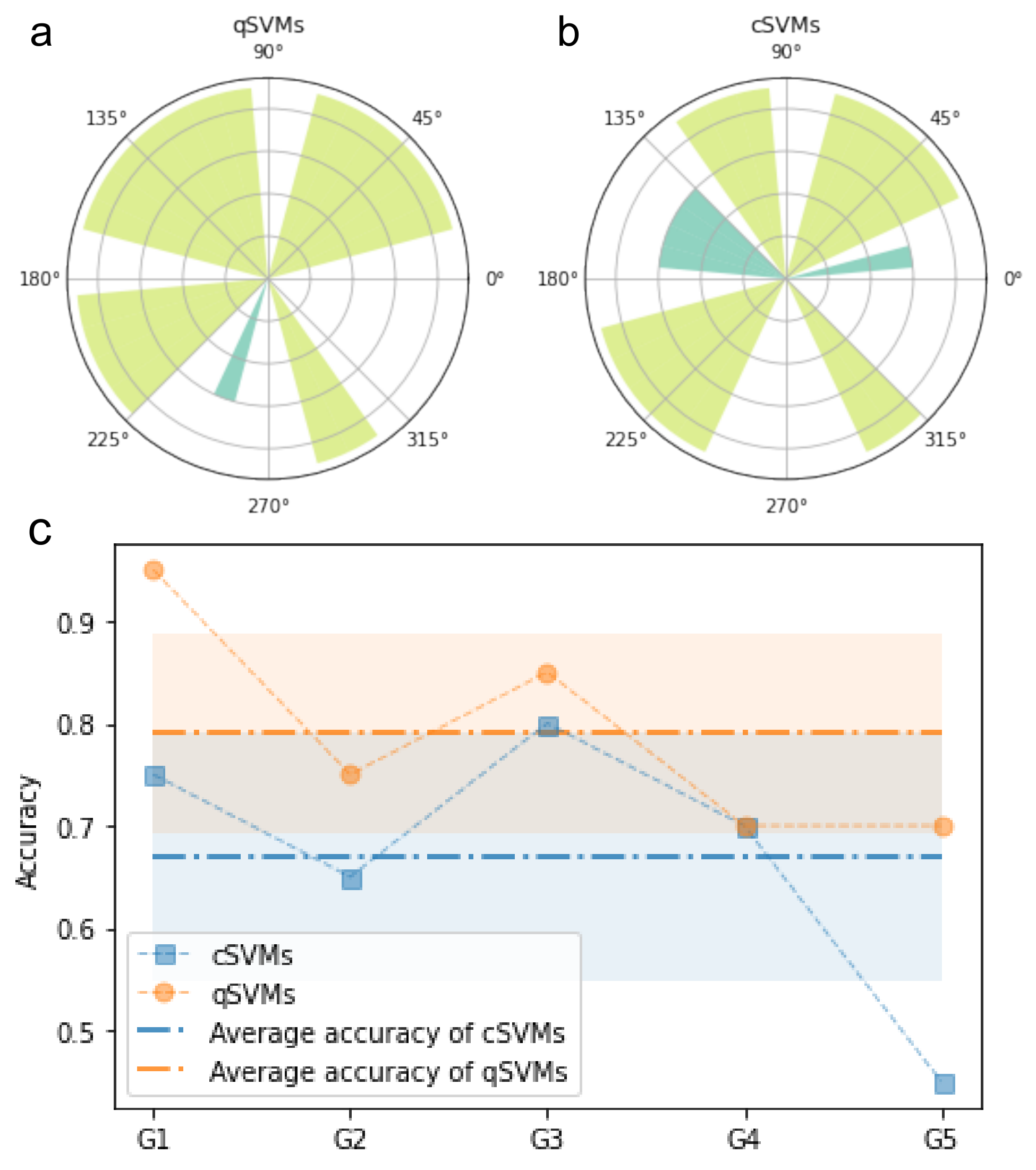} 

    \caption{\quad Results for attacking angle classification. (a-b) Result of the classification from the shuffled dataset G1 using (a) qSVM and (b) cSVM, respectively. Each quarter circle (0-90$^{\circ}$, 90$^{\circ}$-180$^{\circ}$, 180$^{\circ}$-270$^{\circ}$ and 270$^{\circ}$-360$^{\circ}$) corresponds to AOA of 14$^{\circ}$-15$^{\circ}$, 16$^{\circ}$-17 $^{\circ}$, 18$^{\circ}$-19$^{\circ}$ and 20$^{\circ}$ respectively. The sectors represent data from the test set, every 10$^{\circ}$ of the sector corresponding to one point in the test data. The colors of the sectors represent the classification results: light green represents the correct classification, and blue-green represents misclassification into adjacent classes. There is no error that separates two or more classes from the true class. (c) Performance of qSVMs (blue line) and cSVMs (orange red line) as measured by accuracy. The classical multi-classifier we use here is Nusvc from scikit-learn, and the hyperparameters are $\gamma$ determined by "scale" and $nu=0.3$. Each point on the line is the test result of different shuffled data, labeled with G1, G2 et al. The hyperparameters for qSVMs are $B=4, K= 3, \xi = 0, \gamma = 0.27,$ and $\  C = 21$. The average of the classification accuracy is also shown in the figure, 0.67 for cSVMs and 0.79 for qSVMs.  Shaded areas in the figure represent within one standard deviation, where the standard deviation is 0.121 for cSVMs and 0.097 for qSVMs.}
    \end{figure}

\par It is worth to note that for qSVMs, the class of 20$^{\circ}$ data is prone to classification errors. For instance, the blue-green sector in Fig.4a is mis-classified to 20$^{\circ}$, and similar scenarios happen in the cases of G2 and G3 shown in Appendix D. Considering that there are fewer samples in the 20$^{\circ}$ class than other classes, we suggest the unbalanced samples may add some challenges to qSVMs, possibly due to a demand on a suitable setting of bias $b$. We give more detailed explanation in Appendix E, and show that we can improve the the performance for qSVM by adjusting and seeking for a bias $b$ that is suitable for the dataset. 
    
 \section{Discussion and Conclusion}   
In conclusion, we applied the qSVMs to classification problems in fluid dynamics and demonstrated that qSVMs outperformed the classical classifiers for both the binary and the multi-class classification tasks. In detecting whether there is fluid separation, we exhibited that solving QP problem through quantum annealing can effectively realize the function of SVM. By introducing one-against-all algorithm, We built a multi-class qSVM and obtained better accuracy than the classical SVMs on the angle classification problem. The advantage originates from different solutions of the QP problem acquired by D-Wave quantum annealer and cSVM. Using several close-to-optimal solutions instead of the global optimum makes it possible to better generalize the distribution of different classes of points, and by using the value of the decision function to judge the class, the advantage is further amplified. 

To improve the performance of qSVMs, there are several directions to try: the encoding method, the selection of bias $b$, and the embedding process. Since $\alpha _n$ can only take discrete $2^{K}$ values according to Eq. (A3), the appropriate encoding method has to be chosen according to the data characteristics so that the discrete variable solution $\{\alpha_n\}$ is as close as possible to the continuous variable solution $\{\alpha\}$. The difference between the solution obtained by the D-Wave system and the theoretical optimal solution leads to the adjustment of the bias selection. As for the embedding, it is a crucial aspect of mapping the problem under the D-Wave configuration. Note that the update of the D-Wave system makes the QPU structures alter, which in turn brings the update of the embedding, and the performance of the qSVMs will be varied by choosing different D-Wave solvers\cite{willsch2021benchmarking}. The qSVM may further benefit from various recent progresses on quantum annealing techniques such as shortcuts to adiabaticity\cite{Hegade2021}.  

\par In addition, we have to admit that when faced with a large number of classification problems, the one-against-all algorithm will significantly reduce its classification ability due to the imbalance of the samples, and the multi-classifier based on the one-against-one algorithm will play a better role. However, the one-to-one algorithm often takes more time to process due to the increased number of classifiers, so the multi-classifier should be smartly selected according to the problem.

\par To sum up, a data-driven classification method is used to detect flow separation and airfoil stall, which is beneficial for aircraft design and flow control. Inspired by the initial successful application, the quantum computing methods can also be used in real-time detection of aerodynamic force in further study. In real-time flow control, only scattered measurement data is available, which is far from enough to accurately obtain the details of the flow field distribution on the aircraft and the aerodynamic force, i.e. lift and drag. Existing methods of pressure field reconstruction and aerodynamic force detection, i.e. neural network and numerical simulation, usually suffer from heavy data size and long computational time. In this regard, the quantum computing method can possibly provide a faster and more accurate solution. Apparently, there are few quantum computing methods used for on-line flow measurement and control in the past. Our proposed method may give assistance to on-line flow classification and control with other flow feature extraction methods, such as compressed sensing and proper orthogonal decomposition reduced-order model. This is exhilarating for flow real-time control with a wide range of application. With
the enhancement of the computing power of quantum processors, the combination of quantum computing and various disciplines will become more prevalent, and the quantum advantages will be widely reflected.

\newpage

\begin{appendix}
\clearpage
\newpage
\onecolumngrid

\setcounter{table}{0}
\setcounter{equation}{0}
\setcounter{figure}{0}
\setcounter{section}{0}

\renewcommand{\thetable}{{A}\arabic{table}}

\bigskip
\section{Appendix A: Calculation for QUBO }

\renewcommand{\theequation}{{A}\arabic{equation}}
\renewcommand{\thefigure}{{A}\arabic{figure}}

 QUBO, an acronym for a Quadratic Unconstrained Binary Optimization problem, is a mathematical formulation that covers various Combinatorial Optimization problems found in finance, economics and machine learning. QUBO is an NP hard problem and is challenging for current algorithms to solve. However, with some specific embeddings that have already been applied to many classical problems (like Maximum cut, Graph coloring and the Partition problem), QUBO solvers can be used to solve some difficult optimization problems efficiently once they are put into the QUBO framework.
 \par Both Ising model and QUBO have the binary quadratic formulation and they can be easily converted to each other. The Ising model is a kind of stochastic process model describing the phase transition of matter and is traditionally used in statistical mechanics. It uses the spins as variables, “spin up” (↑) and “spin down” (↓) states correspond to +1 and -1 values, respectively. Relationships between the spins are represented by couplings. The objective function of the Ising model is expressed as follows:
 \begin{equation}
     E_{Ising(s)} = \sum_{i=1}^{N}h_i s_i +\sum_{i=1}^{N}\sum_{j=i+1}^{N}J_{i,j}s_is_j,
 \end{equation}
 where $h_i$ is the linear coefficient corresponding to the qubit biase, and $J_{i,j}$ is the quadratic coefficient that represents the  coupling strength.
\par The QUBO is defined by the optimization problem:
\begin{equation}
    minimize \qquad x^T Qx,
\end{equation}
where $x\in \{0,1\}^n$ is a vector of binary variables and $Q$ is a $n \times n$ square matrix of constants. In general, the Q matrix is described as symmetric or upper triangular forms, without loss of generality: 
\par Symmetric form: For all $i$ and $j$ except $i = j$, replace $q_{ij}$ by ($q_{ij}+q_{ji}$) /2.
\par Upper triangular form: For all $i$ and $j$ with $j>i$ replace $q_{ij}$ by ($q_{ij}+q_{ji}$). Then replace all $q_{ij}$ for $j<i$ by 0. If the matrix is already symmetric, this just doubles the $q_{ij}$ values above the main diagonal, and then sets all values below the main diagonal to 0.
\par Due to this structural similarity QUBO offers an opportunity for the application of quantum annealing, since the hardware foundation of D-Wave's quantum annealing algorithm is built by Ising model. Because of the connection, QUBO has become one of the central problems of quantum computing and research. 
\par For the QUBO used in our research, we took two steps to construct it for use on D-Wave quantum solver. In the first step, the problem was formulated as an objective function. To formulate the QP problem given in Eqs. (2)-(4), an encoding was used to adopt binary variables.
\begin{equation}
    \alpha_n=\sum_{k=0}^{K-1}B^{k}a_{Kn+k},
\end{equation}
where $a_{Kn+k}\in \{0,1\}$ are binary variables, $K$ is the number of binary variables to encode $\alpha_n$, and $B$ is the base used for the encoding. On the basis of our parameter selection process, we found that $K=2 \ or \ 3$ and some small $B$ values are sufficient to get reasonable results. Through encoding, the QUBO form can effortlessly satisfies constraint in Eq. (3) by defining 
\begin{equation}
    C=\sum_{k=1}^{K}B^{k}.
\end{equation}
\par After encoding, we introduce a multiplier $\xi$ to include the second constraint given in Eq. (4) as a squared penalty term, and the QUBO form can be obtained.
\begin{equation}
\begin{split}
        E=\frac{1}{2}\sum_{nmkj}a_{Kn+k}a_{Km+j}B^{k+j}t_nt_mk(x_m,x_n) \\ -\sum_{nk}B^{K}a_{Kn+k}+\frac{\xi}{2}\Big(\sum_{nk}B^{K}a_{Kn+k}t_n\Big)^2
\end{split}
\end{equation}
\begin{equation}
        =\sum_{n,m=0}^{N-1}\sum_{k,j=0}^{N-1}a_{Kn+k}\widetilde{Q}_{Kn+k,Km+j}a_{Km+j},
\end{equation}
where $\widetilde{Q}$ is a symmetric $KN \times KN$ matrix determined by 
\begin{equation}
\begin{split}
        \widetilde{Q}_{Kn+k,Km+j}=&\frac{1}{2}B^{k+j}t_nt_m(k(x_n,x_m)+\xi)\\&-\delta_{nm}\delta_{kj}B^{k}.
\end{split}
\end{equation}
Since the D-Wave system uses the upper triangular form to solve the QUBO problem, $\widetilde{Q}$ is substituted by $Q_{ij}=\widetilde{Q}_{ij}+\widetilde{Q}_{ji}$ for $i<j$ and $Q_{ii}=\widetilde{Q}_{ii}$. 
\par The next step is to select a quantum solver and execute the task via the D-Wave system. In our work, we connected to the Advantage\_system4.1 in North American sever. To execute the task, there still remains an essential process called embedding which aims at increasing logical connections between qubits. Because the objective function can be represented as a graph, it can be mapped to the QPU (linear coefficients to qubit biases and quadratic coefficients to coupler strengths). The QPU uses quantum annealing to seek for the minimum of the resulting energy landscape, which corresponds to the solution. Since the connection within the QPU in D-Wave Advantage\_system is given by the Pegasus topology, it is not guaranteed that all qubits are well connected to each other, and through embedding the logical qubits are introduced and the connection is correspondingly enhanced. When the embedding process fails, one should gradually reduce the number of nonzero couplers $n_{cpl}$ until the embedding is found. D-Wave Ocean SDK provides a module named Embedding Composite and can automatically generate qubit embeddings.

\section{Appendix B: Experimental Setup of the real data  }
The data was provided to us on request. The experiment was performed in a $1000 mm$ long open-circuit wind tunnel with a cross section of $300\times300 mm^{2}$. The airfoil model $NACA0018$ was fixed at the middle height of the test section. The average freestream velocity $U_{\infty}$ was maintained at $10$, $13$, and $17 m/s$ to study the characteristics of flow at different Reynolds numbers. The angle of attack was adjusted in the range $\alpha = 0-19 $ $^{\circ}$ by turning the rotary table. When $\alpha \le 10$ $^{\circ}$, the resolution of $\alpha$ is 2 $^{\circ}$, and when $\alpha  \textgreater 10$, the resolution of $\alpha$ is 1 $^{\circ}$. In this way, the variation of $\alpha$ results in 15 cases under every freestream velocity.

\par A chordal separation pressure tap is arranged on the upper wing surface of the model to accurately detect the separation. Totally 10 pressure taps are distributed nonlinearly on the section at the midspan of the airfoil model. The differential pressure was measured during the experiment and used for calculation. More details about the experimental setup are provided in Zhou et al\cite{Zhou2021}.
\par The CFD simulation is based on the airfoil $NACA63_{3}-018$ , the blue outline shown in Fig.A1 represents the airfoil profile. A total of 106 pressure measuring points marked with a red circle are evenly distributed on the whole wing surface, and 53 pressure measuring points are set on the upper and lower wing surfaces respectively. The angle of attack was adjusted in the range $\alpha = 0-20 $ $^{\circ}$ successively, where the resolution of $\alpha$ is 1 $^{\circ}$. The freestream velocity ranged from $40 m/s$ to $120 m/s$, where the resolution is $10 m/s$. As such, we had 189 simulated cases in total to study the characteristics of the flow. 

\par ANSYS Fluent is used as the solver for 2D simulation calculation. The calculated airfoil chord length is 450mm, and the calculated external field size is set to 10 times of the chord length. The grid adopts O-shaped structured grid. The length of the first layer of grid is calculated according to the and $y+$ conditions, and the grid at the leading edge of the airfoil is encrypted. The grid independence is verified before calculation. In the setting of solution conditions, the turbulence model is $SST\;k-\omega$. the turbulence degree is taken as $2\%$. The Sutherland equation is selected as the viscosity equation. The calculation is based on the implicit algorithm with density basis. All spatial resolutions are set as second-order difference. The whole simulation calculation is completed by the workstation of the Supercomputing Center of Shanghai Jiao Tong University.

\begin{figure}[H]
	\centering
	\includegraphics[scale = 0.5]{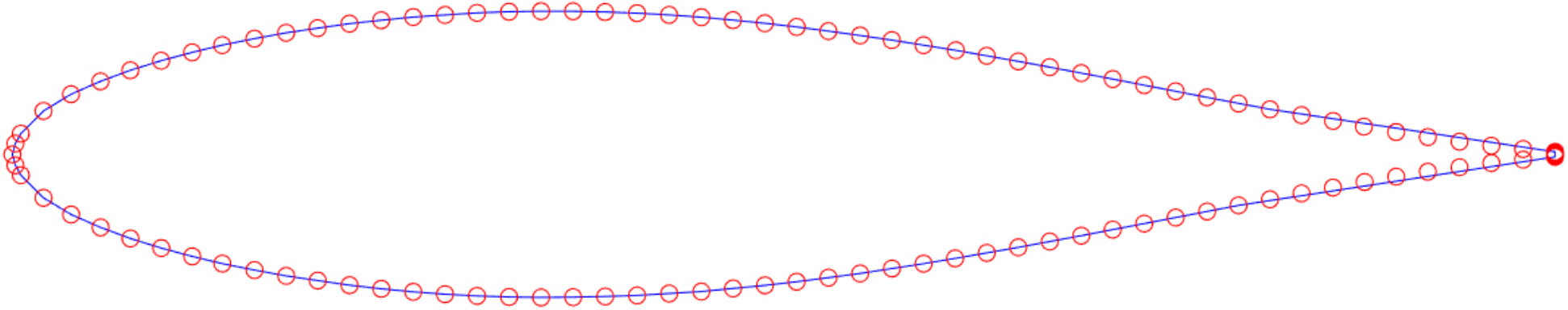}
	\caption{Schematic diagram of airfoil $NACA63_{3}-018$ and simulated pressure measuring point.}
\end{figure}

\section{Appendix C: Indicators of binary classification  }
For binary classification problems, there are several indicators to comprehensively evaluate the classification: Accuracy = (TP+TN)/(TP+TN+FP+FN) indicates the proportion of correctly predicted samples in all samples, Precision = TP/(TP+FP) indicates the proportion of correctly predicted samples in predicted positive samples, Recall = TP/(TP+FN) indicates the number of correctly predicted data in those data whose real category is positive$t_n=1$, F1-score = 2$\cdot$ Precision $\cdot$ Recall/(Precision+Recall) indicates the harmonic mean of Precision and Recall. All indicators are derived from Confusion Matrix, which is the most basic and intuitive method to measure the accuracy of classification models. It can effectively visualize the accuracy of the classification model corresponding to each category, especially when the number of classification data is unbalanced. The Confusion Matrix has the following form:

\begin{table}[!ht]
    \centering
     \caption{Confusion Matrix. The columns of the confusion matrix represent the predicted classes and the rows represent the actual classes.The values on the diagonal represent the number or ratio of correct predictions, the off-diagonal elements are the parts that were predicted incorrectly.Therefore the diagonal values of the confusion matrix are expected to be as high as possible, indicating more correct predictions.}
    \begin{tabular}{|l|c|c|}
    \hline
    \diagbox{Real class}{Prediction class} & $\widetilde{t_n}=1$ & $\widetilde{t_n}=-1$  \\
    \hline
    $t_n=1$ & True Positive (TP) & False negateive (FN) \\
    \hline
    $t_n=-1$ & False Positive (FP) & True negative (TN) \\   
    \hline
    \end{tabular}
   
\end{table}

\begin{figure}[b]
\includegraphics[scale=0.37]{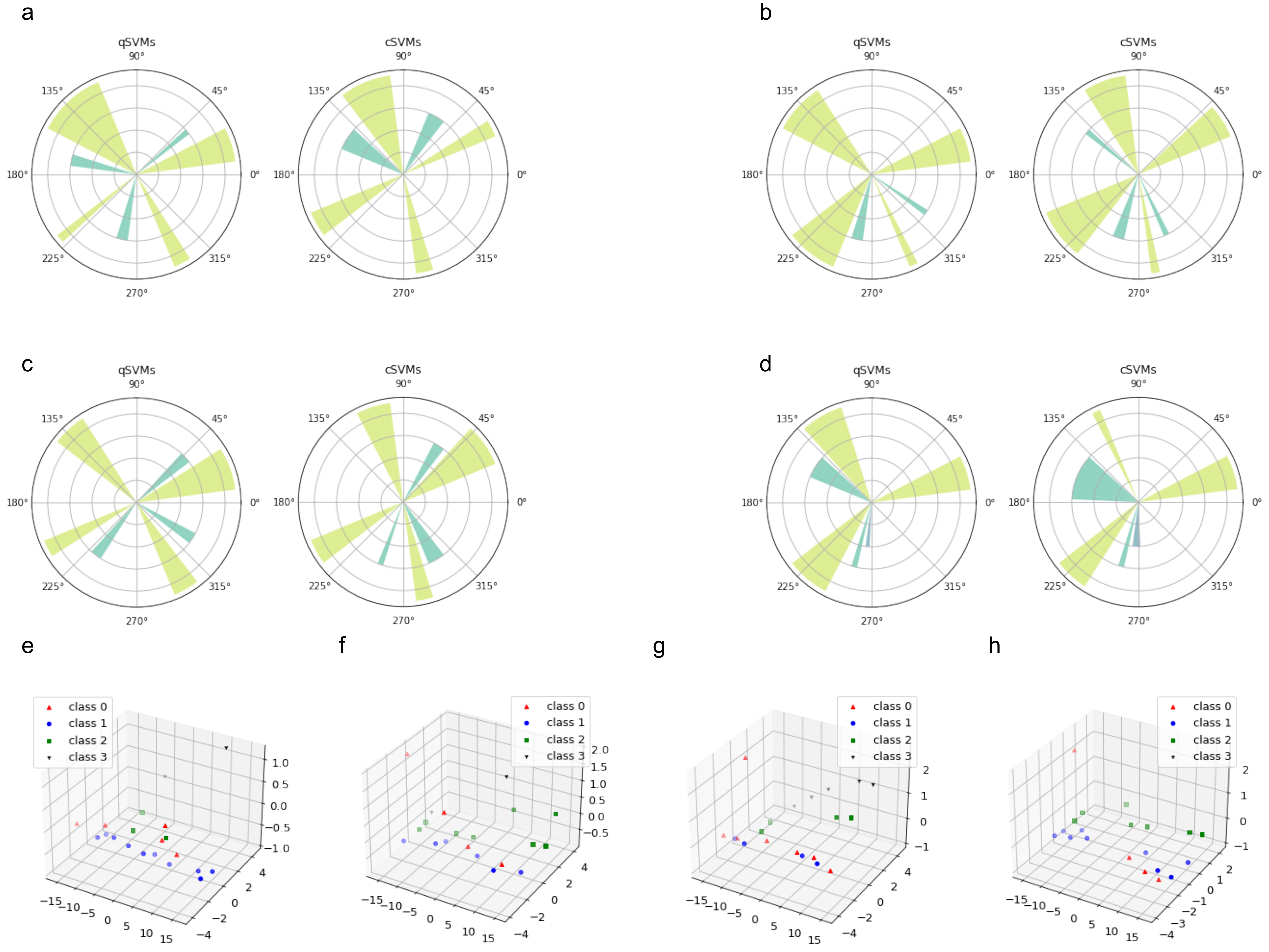}
\caption{Data and results of AOA classification. (a)-(d) correspond to the classification results of the G2-G5 dataset in the main text. Again, the width of the sector represents the amount of data, the sector with the largest radius represents the correctly classified data, the middle sector represents the data classified to the adjacent class, and the sector with the smallest radius represents the classification results to the wrong category other than the adjacent class.(e)-(h) represent the distribution of the training set of G2-G5 under the 3D feature space}
\end{figure}  

\section{Appendix D: Result of AOA classification  }
In this part we present the data shuffled in different orders in the AOA classification and the corresponding classification results. As described in the main text on the process of handling AOA classification, the dataset is obtained from CFD simulation. For the data before the flow field separation, we found that the classification accuracy of different classifiers is very high, hence the classification is not difficult and its result is not displayed in the article. We use the actual data after flow separation with AOA from 14$^{\circ}$ to 20$^{\circ}$, where $D=\{(x \in R^{106}, t_n = 0,1,2,3),n=0.\cdots,62\}$. After PCA processing, the dataset is reduced to three dimensions to reduce the computational overload. The distribution of the obtained Training set in 3D feature space is shown in FIG.A2(e)-(h). The more mixed data points means the more difficult to classify and the classification results by different classifiers are shown in Fig.A2(a)-(d).

\section{Appendix E: Adjusting the bias for qSVMs}
As the solutions returned by the quantum annealer are some close-to-optimal solutions, which are not strictly equal to the global minimum $\{ \alpha_n\}$, it is necessary to adjust the bias determined by Eq.(6). We found that this deviation is obvious in unbalanced samples, so we adjusted the bias of qSVM for distinguishing 20$^{\circ}$, and improved the classification ability of qSVMs.

Since other parameters are not constrained by bias $b$, $b$ can be adjusted independently without requiring a new training of qSVM, and we finally get the $b^*$ that makes the training set the most accurate. We select it in the range of one near bias, and find $b^* = b +0.5$ to effectively improve the accuracy of training set. The adjusted classification results of qSVMs are shown in Fig. A3.
\begin{figure}[H]
    \centering
    \includegraphics[scale = 0.7]{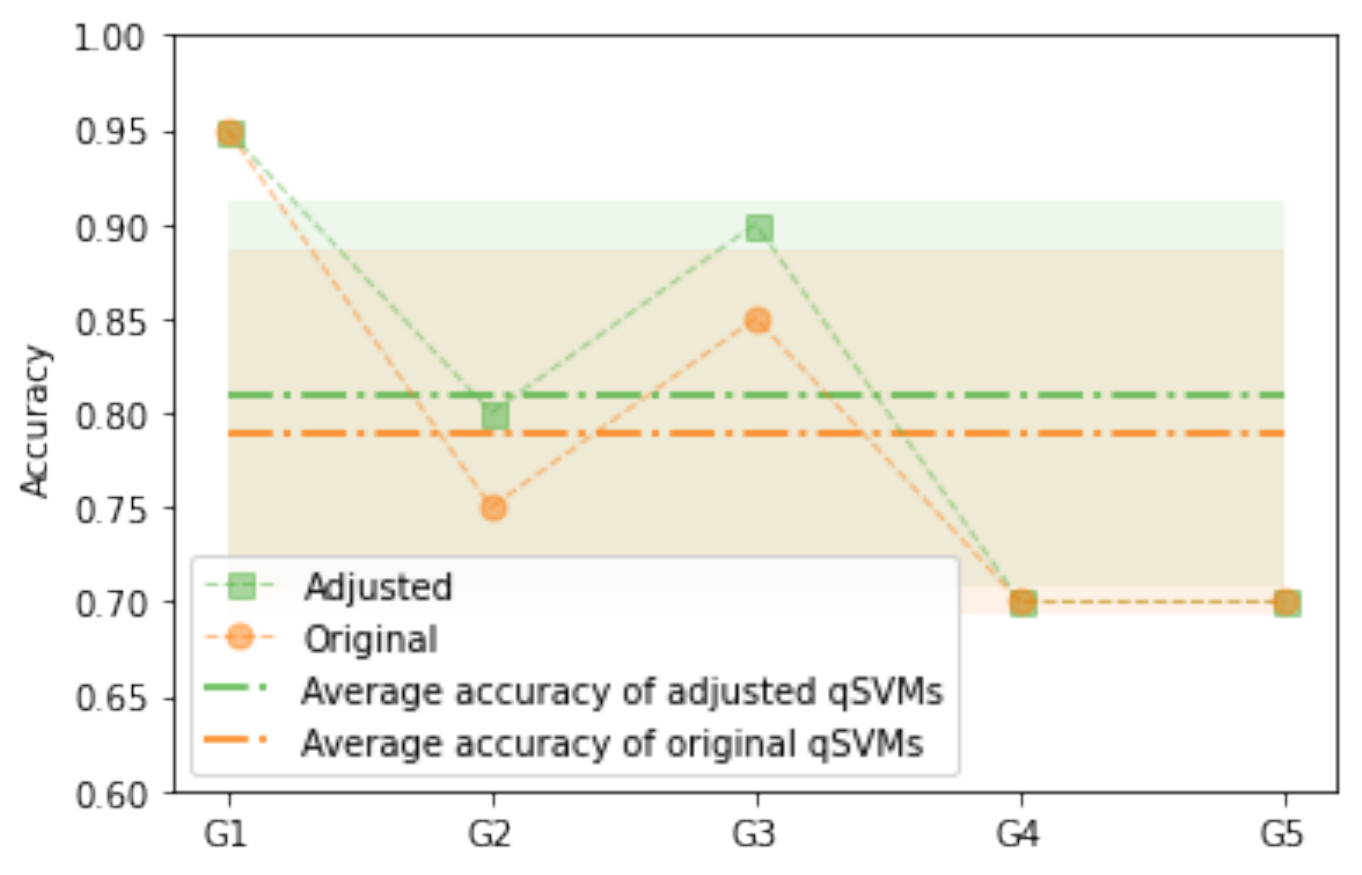}
    \caption{Classification accuracy of qSVMs after adjusting the bias. $b =-0.26$ and $b^* = 0.24$. After adjusting the bias, the accuracy of qSVMs is improved to 0.81 and the standard deviation is 0.708. Other parameters remain unchanged.}
\end{figure}

\end{appendix}
\bigskip
\end{document}